\documentclass[preprint,aps,pdf]{revtex4}
\usepackage{amsmath}
\usepackage{amssymb}
\usepackage{epsfig}
\usepackage{graphicx}
\usepackage{amsfonts}
\usepackage[usenames]{color}

\newcommand{\de}{\textrm{d}}

\newcommand{\tpr}{t^{\prime}}

\newcommand{\dco}{c_1^{(\omega)}}

\newcommand{\ofb}{f^L_{\beta}}
\newcommand{\ogb}{g^L_{\beta}}

\begin{document}

\title{Anomalous non-linear response of glassy liquids: general arguments and a Mode-Coupling approach} 

\author{Marco Tarzia}
\affiliation{Institut de Physique Th{\'e}orique,
CEA, IPhT, F-91191 Gif-sur-Yvette, France
CNRS, URA 2306, F-91191 Gif-sur-Yvette, France}
\author{Giulio Biroli}
\affiliation{Institut de Physique Th{\'e}orique,
CEA, IPhT, F-91191 Gif-sur-Yvette, France
CNRS, URA 2306, F-91191 Gif-sur-Yvette, France}
\author{Jean-Philippe
  Bouchaud}
\affiliation{Science \& Finance, Capital Fund Management, 6 Bd
Haussmann, 75009 Paris, France}
\affiliation{Service de Physique de l'{\'E}tat Condens{\'e},
Orme des Merisiers -- CEA Saclay, 91191 Gif sur Yvette Cedex, France}
\author{Alexandre Lef{\`e}vre}
\affiliation{Institut de Physique Th{\'e}orique,
CEA, IPhT, F-91191 Gif-sur-Yvette, France
CNRS, URA 2306, F-91191 Gif-sur-Yvette, France}
\begin{abstract}
We study theoretically the non-linear response properties of glass formers. We 
establish several general results which, together with the assumption of
Time-Temperature Superposition, lead to a relation between the non-linear response 
and the derivative of the linear response with respect to temperature. Using results from 
Mode-Coupling Theory (MCT) and scaling arguments valid close to the glass transition, we obtain 
the frequency and temperature dependence of the non-linear response in the $\alpha$ and $\beta$-regimes. 
Our results demonstrate that supercooled liquids are characterized by responses to external perturbations that become increasingly non-linear as the glass transition is approached.
These results are extended to the case of inhomogeneous perturbing fields.
\end{abstract}
\maketitle


\section{Introduction and motivations}\label{sec:intro}
 
Physical systems become
very susceptible to external perturbations close to a phase transition. 
In several theoretical approaches the huge increase of the relaxation time of 
super-cooled liquid is traced back to the proximity to a phase 
transition~\footnote{Whether the transition is avoided or it takes place at finite or zero temperature depends
on the approach}. Therefore, it would be natural to expect that responses to external fields become singular also 
approaching the glass transition. However, standard {\it linear} responses (as well as correlation functions)
are known to remain quite featureless. In this work, following up~\cite{BB-PRB}, we show 
using general arguments that instead {\it non-linear} responses do increase. 
Furthermore, we work out detailed predictions within
the Mode Coupling Theory of glasses and we show what is the relation between the growth of 
non-linear responses and the one of previously introduced probes of dynamic
correlations.    
      
In order to grasp why only non-linear responses may grow 
approaching the glass transition, it is instructive to recall 
the situation of standard critical phenomena and contrast it to the 
one of spin glasses. In standard critical phenomena 
there is a spontaneous symmetry breaking toward an ordered
phase. The main consequence of the broken symmetry is that 
there are different but thermodynamically equivalent states 
in which the system can be found: they are all related by the 
symmetry at hand. For example, below the ferromagnetic transition
the state with a given magnetization $M$ and the one characterized
by the magnetization $-M$ are equivalent and the system will be in one 
or the other depending on boundary conditions, residual fields, etc.  
Switching on an external field that couples to the order parameter allows one
to select one given state, e.g. a positive magnetic field selects the state with positive
magnetization in ferromagnetic systems. As a consequence, it is easy to understand
that just before the transition, where the system is at the brink of developing
long range order, the response to an external field is huge and actually becomes
infinite at the critical point. The situation is more subtle for systems 
characterized by the appearance and growth of {\it amorphous long range order}, 
as it is the case in spin glasses and is conjectured to be the case for glasses~\cite{Wolynes,BB-JCP,Nature}. 
The only diverging {\it linear} susceptibility would be the {\it staggered} one: 
the response obtained by using a small field with spatial modulation correlated to the one of the amorphous state. Of course, 
this is not doable in experiments and not even in simulations since it is impossible to guess the external field which would impose 
a given {\it amorphous} long-range ordered state because it is amorphous as the state it pins (an external field with an uncorrelated spatial modulation, e.g. an homogenous field, 
does not allow to select one state from the others and so it does not lead to singular responses). 
For spin glasses, the square of the staggered linear response can nevertheless be probed,  since it is related to the static third order  
non-linear susceptibility~\cite{BY,LLevy,DHuse} , which was shown to diverge at the transition. Similar arguments~\cite{BB-PRB} (see also below) 
adapted to supercooled liquids - where the disorder is not quenched but self-induced - suggest that linear responses will be featureless
but non-linear susceptibilities should instead increase approaching the glass 
transition. The main difference between spin glasses and structural glasses
is that in the latter case one has to focus on {\it dynamical} 
non-linear responses. It was argued in ~\cite{BB-PRB}, on the basis 
of physical and heuristic arguments, that the non-linear dielectric susceptibility $\chi_3(\omega)$ should exhibit a 
growing peak around $\omega \tau_\alpha = 1$, while $\chi_3(\omega = 0)$ should remain trivial, in contrast with the case of spin-glasses 
($\tau_\alpha$ is the standard notation for the relaxation time of super-cooled liquids). 
However, the detailed shape of $\chi_3(\omega)$ in the glassy region is beyond the grasp of those heuristic arguments. Since the 
corresponding experiments are currently being performed ~\cite{SaclayGroup,Richert}, it is quite important to get more precise predictions 
on the expected shape of $\chi_3(\omega)$. This is the primary aim of the present study, where we obtain for the first time, using general 
arguments made more precise in the context of the Mode Coupling Theory (MCT)~\cite{Gotze} of the glass transition, some precise information on
the non-linear susceptibility, in particular concerning both its frequency and temperature dependence. 
Our results can be generalized to other non-linear responses such as non-linear 
compressibility and non-linear rheological responses. These should be measurable for colloids
approaching the glass transition.
 
It is important to relate the results on non-linear responses to the 
ones obtained recently on dynamical correlations. This is the second aim of our work. 
For purpose of self-consistency, we recall below some definitions and results that we will use
in the following sections. 

It was established in the last decade that glass-forming liquids become
more and more dynamically correlated approaching the glass transition~\cite{reviewDH}. 
This phenomenon, related to an increasing heterogeneity in the dynamics, is remarkable since, for the 
first time, some type of spatial correlation has been clearly connected to the slowing down of the dynamics of supercooled liquids. 
In order to unveil the existence of dynamical correlations, one has to focus on some local 
probe of the relaxation dynamics - typically a two point function $\mathcal{O}(x,t)= \phi(x,t)\phi(x,0)$
where $\phi(x,t)$ could be the density fluctuation, $\delta\rho(x,t)$, at position $x$ and time $t$, or the mobility
field~\cite{GC}.  A measure of dynamical spatial correlations is obtained by considering the four point correlator 
$G_4(x-y;t)=\langle \mathcal{O}(x,t) \mathcal{O}(y,t)\rangle_c$~\cite{oldIndians,FP}, where the brakets
denotes the usual connected average over dynamical histories. 
By analogy with standard critical phenomena, a 'susceptibility' \footnote{For some technical reasons the 
function $\chi_4$ was called 'susceptibility' in~\cite{FPG}. This terminology is still used
although is a measure of the fluctuation of the dynamics and not a response function.} may be naturally defined by spatially integrating $G_4$: $\chi_4(t)=\int dx\, G_4(x,t)$. 
This `dynamical susceptibility', $\chi_4$, has been intensively studied in the past few years, both 
theoretically and numerically~\cite{FPG,Lacevic,Berthier,TWBBB,JCP1,JCP2,epl}. 
One finds that $\chi_4$ reaches a peak value for time scales of the order
of the relaxation time of the system $\tau_\alpha$, and the height of this peak increases as the 
temperature is reduced, as a clear sign of the growth of some dynamical correlation length as the 
glass transition is approached. From an experimental point of view, however, four-point correlation
functions are very difficult to measure directly, except in cases where one can monitor the
trajectory of individual particles - for example granular and colloidal systems where $\chi_4$ can be measured directly
and again shows interesting features as the system jams~\cite{Dauchot1,Doug,Dauchot2,Luca}.  

Another interesting quantity, extensively studied in the context of dynamic heterogeneity in the past few years, is the derivative of the standard two-body correlation 
$C(\tau)$ (or susceptibility $\chi_1(\omega)$) with respect to the temperature (or the density) - 
a quantity called $\chi_T=T\partial C(\tau)/\partial T$ (or $\chi_\rho=\rho\partial C(\tau)/\partial \rho$) in~\cite{science}. 
This is a non-standard linear dynamical response and is 
clearly an easily accessible quantity, which also shows a peak at times of the order of $\tau_\alpha$  
and whose height grows as the temperature is lowered~\cite{science,PRE}. This has lead to direct estimates of 
the size of the dynamical correlation length in supercooled liquids and glasses~\cite{science,PRE,Ruocco}. The relation
between $\chi_4$ and $\chi_T$ is however highly non-trivial and has been investigated thoroughly in~\cite{JCP1,JCP2}. 
It was realized in these papers that the existence of conserved quantities (energy, density) crucially affects the
properties of $\chi_4$, which depends both on the thermodynamic ensemble (NVE vs. NPT for example) and on the
dynamics (Brownian vs. Newtonian for example). The true glassy correlation length, on the other hand, does not depend 
on these choices, and therefore the direct interpretation of $\chi_4(t)$ in terms of a correlation volume is somewhat obscured. 
At a deeper level, the basic ingredient leading to the critical behaviour of $\chi_4$ turns out to be entirely contained in the response 
function $\chi_T$ itself, as the field theoretical analysis of~\cite{JCP1} explicitely demonstrates and the numerical results presented there
fully confirm. For example, for Brownian dynamics or for Newtonian dynamics in the NVE ensemble, $\chi_4 \approx \chi_T$, whereas 
for Newtonian dynamics in the NVT ensemble, one rather finds $\chi_4 \approx \chi_T^2$, see~\cite{JCP1,JCP2} for a detailed discussion.

\subsection{Summary of results} 
\begin{figure}
\begin{center}
\includegraphics[width=0.74\textwidth]{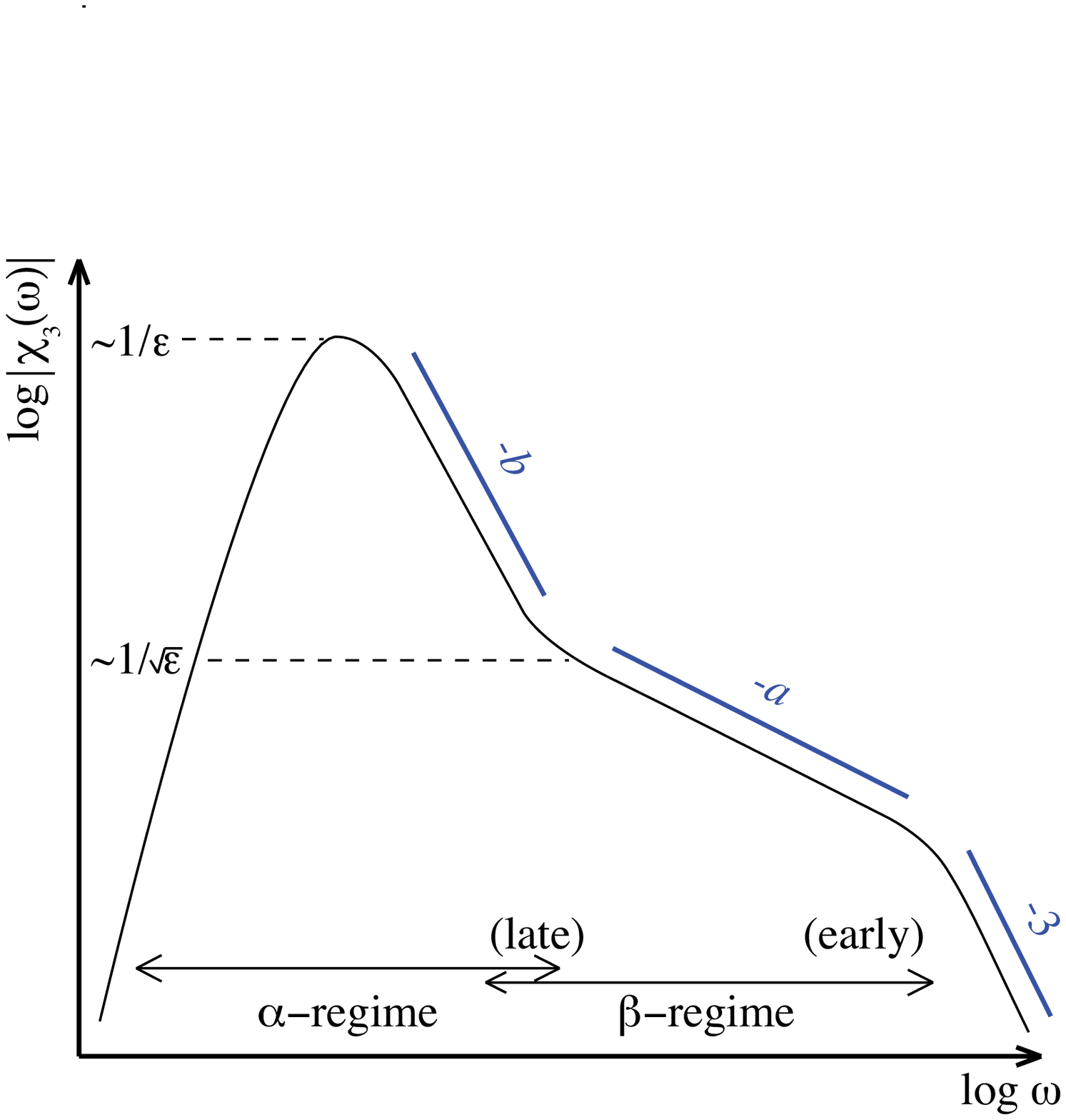}
\end{center}
\caption{Sketch of $\log |\chi_3 (\omega)|$ as a function of
$\log \omega$, showing five different frequency regimes: $\omega
\tau_\alpha\ll 1$, $\omega \tau_\alpha \sim 1$,
$\tau_\beta/\tau_\alpha\ll\omega \tau_\beta \ll 1$ ($\epsilon=T-T_c$), $\omega \tau_\beta \gg 1$, $\omega \tau_0 \sim 1$.
Note that the low frequency limit is non zero but much smaller than the peak value for $T$ close to $T_c$.}
\label{fig:chi3}
\vspace{0.0cm}
\end{figure}

One of the two main results of this work consists in obtaining the precise relation between 
non-linear responses and probes of dynamical correlations introduced previously in the literature. In particular,
we shall focus on the third-order non-linear dielectric response  $\chi_3(\omega)$ and work out its 
relationship with the dielectric dynamical susceptibility $\chi_T$. The latter is known to be related to 
$\chi_4$, see~\cite{JCP1} for a detailed discussion. 
We will establish simple identities between $\chi_3$ and $\chi_T$ which hold whenever 
Time-Temperature Superposition (TTS) holds, i.e. whenever all dependence of the linear response of the system on external parameters (temperature,
density, electric field, \ldots) comes through the dependence of the
relaxation time on these parameters. This is a strict statement within
MCT, where a true dynamical phase transition takes place at a finite
temperature $T_c$~\cite{Gotze}. More precisely, we will show that at low enough frequency (in fact much smaller than the inverse of the relaxation time), the following relation holds:
\begin{equation}\label{eqn:rel}
\Im\left(\chi_3(\omega)\right)\approx\frac{\kappa\omega}{T^2}\widehat\chi_T(2\omega), 
\end{equation}
where $\kappa$ is a slow varying function of the temperature, see the following sections for a precise definition.
This relation is actually a consequence of a more general one relating $\chi_3(\omega)$ to the standard linear response $\chi_1(\omega)$,
which is also valid only at low frequency and reads: 
\begin{equation} \label{eqn:rel2}
\chi_3(\omega) \approx  \kappa \frac{\partial \chi_1(2 \omega)}{\partial T},
\end{equation}
These simple relations do not extend over the whole frequency domain, however one could argue that
they should do so in a scaling sense (as long as the TTS is valid). A general proof based on field theory along the lines of~\cite{JCP1} is left for the future. 
In this work, we have checked that within MCT this is indeed
what happens. Furthermore, within MCT, we have also worked out the complete critical behavior of 
$\chi_3$, which is sketched in Fig.~\ref{fig:chi3}. 

\begin{itemize}
\item In the $\alpha$-regime, i.e. $\omega\sim 1/\tau_\alpha\sim\epsilon^{1/2a+1/2b}/\tau_0$, $\chi_3(\omega)$ growths and reaches its maximum, of height of order $1/\epsilon$, after which it decreases as $\omega^{-b}\tau_\beta$ at large $\omega$. In this regime, one has the scaling form: $\chi_3(\omega)=\frac{1}{\epsilon}{\cal G}(\omega\tau_\alpha)$.
\item At the crossover between the early $\alpha$-regime and late $\beta$-regime $\chi_3(\omega)$ is of order $1/\sqrt{\epsilon}$.
\item In the $\beta$-regime, i.e. $\omega\sim 1/\tau_\beta\sim\epsilon^{1/2a}/\tau_0$, $\chi_3(\omega)$ dicreases as $\omega^{-b}\tau_\beta$ at small $\omega$ and as $\omega^{-a}\tau_\beta$ at large $\omega$, with scaling form $\chi_3(\omega)=\frac{1}{\sqrt{\epsilon}}{\cal F}(\omega\tau_\beta)$
\end{itemize}
Exponents $a$ and $b$ are well known critical exponents of MCT, which characterize respectively how the correlators  decay and exit of the plateau; $\tau_0$ is a microscopic relaxation time. All along this paper, $\epsilon = T - T_c$, the distance from the Mode Coupling critical temperature $T_c$, will be our control parameter. We remark that the existence of the peak and the decrease at low frequency is a non trivial 
prediction since  it is in contrast to what happens for the (trivial) non-linear response of uncorrelated Brownian dipoles~\cite{perpignan} and for spin glasses (the decrease with an exponent three at high frequency 
sketched in Fig.~\ref{fig:chi3} is instead trivial and will be discussed later). 

For symmetry reasons, the quadratic non-linear susceptibility $\chi_2(\omega)$ is zero for unpolarized systems. This would 
not be the case for example for polarized systems, or when considering
the response to a density perturbation. Indeed, the non-linear 
response to a density perturbation contains a quadratic term. The arguments presented below make it clear that in that case
$\chi_2(\omega)$ itself is directly related to $\chi_T$. We therefore expect that all the results presented below
can be straightforwardly generalized to describe the long-wavelength ($q \to 0$) non-linear compressibility (or more general non-linear rheological responses) of supercooled liquids and colloids. Moreover, by adapting these arguments to the explicit results obtained for 
Inhomogenous Mode Coupling Theory~\cite{IMCTJ} we will also extend these results to finite wavevectors $q$ and find that whenever TTS holds, $\chi_3(\omega)$ takes in the $\alpha$-region the scaling form 
conjectured in~\cite{BB-PRB}:
\begin{equation}
\chi_3 (\omega) \simeq \xi_\alpha^{2 - {\overline \eta}} \, {\cal G} (\omega \tau_\alpha),
\end{equation}
where $\xi_\alpha$ is the dynamical correlation length which also appears in $\chi_T$, and ${\overline \eta}$ a certain critical exponent that within MCT is equal to minus 2.

\subsection{Organisation of the article}

The organisation of the paper is as follows. 
We first introduce the theoretical framework needed to deal with
non-linear responses to an  
external field and establish some general relations between different
quantities that naturally appear (Section~\ref{sec:1}). We then
exploit - when they exist - the  
Time-Temperature superposition (TTS) properties of the correlation
function of glassy systems to establish in Section \ref{sec:2} general relations, valid at low frequency, 
between $\chi_3$ and the temperature derivative of the linear response. 
The critical behavior of $\chi_3(\omega)$ is obtained in Section~\ref{sec:3}
using scaling arguments and within MCT. 
We finally discuss the extension of
these results to spatially inhomogeneous perturbing fields and beyond
MCT. We end by a conclusion 
with open problems, possible extensions and experimental suggestions.

\section{Non-linear susceptibility: general framework} \label{sec:1}

In this Section, we introduce the formalism needed to deal with
non-linear response and  
establish a general relation between the non-linear susceptibility and
a dynamical response function recently 
introduced in the literature, which was argued to capture the critical
spatio-temporal 
correlations of the dynamics in the glassy region. In order to remain
close to recent and  
ongoing experiments on glycerol, we use below the language of
dielectric susceptibility.  
However, as mentioned in the Introduction, our arguments and results
can be extended to more general non-linear susceptibilities (mechanical,  
magnetic, etc.). 

\subsection{Linear and non linear response: small field expansion}

Let us consider a dipolar molecular liquid in presence of a 
small external electric field oscillating at frequency
$\omega$ in the $z$-direction. We denote it as:
\begin{equation} \label{eq:field}
{\mathbf E}(t) = {\mathbf z} E(t) \equiv {\mathbf z} E \, \cos (\omega t),
\end{equation}
where $\mathbf z $ is the unit vector in the $z$ direction and $E(t)=E \, 
\cos (\omega t)$ is its $z$-component with peak field amplitude $E$

When the external field is sufficiently small, the polarization vector (per particle) ${\mathbf P}(t,E)$ can be expanded in powers of $E$. In the following we will denote $P(t,E)$ its $z$-component.
Due to the rotational symmetry in the $x$-$y$ plane the other components 
are identically zero. Furthermore, because of the up-down symmetry 
in the $z$ direction, the polarization must be an odd function of $E$, 
i.e., $P(t,-E)= - P(t,E)$. As a consequence, the expansion of $P$ in powers of $E$ contains only 
odd terms:
\begin{equation} \label{eq:a1}
P(t,E) = P_1(t) E +  P_3(t) E^3 + O(E^5),
\end{equation}
where $P_1(t)$ and $P_3(t)$ can be expressed as functional derivatives of the {polarization} 
with respect to the external field:
\begin{eqnarray} \label{eq:d3p}
P_1(t)  &\equiv& \int_{t_1<t} \de t_1 \, \frac{\delta P(t)}{\delta E (t_1)} \bigg 
|_{E=0} \cos (\omega t_1)\\
\nonumber
P_3(t) &=& \frac{1}{6} \int_{t_1,t_2,t_3 < t} \de t_1 \, \de t_2 \, \de t_3 \, 
\frac{\delta^3 P(t)}{\delta E (t_1) \, \delta E (t_2) \, \delta E(t_3)}
\bigg |_{E=0} \, \cos (\omega t_1)\,\cos (\omega t_2)\,\cos (\omega t_3).
\end{eqnarray}
It is important to remark that the linear and non-linear response kernels
in the above integrals are time translation invariant (TTI), i.e. they do not change
if all time variables are shifted by the same amount. This comes from the 
fact that they are {\it equilibrium} response functions, measured in absence of
the external field. Using this
result and the specific form of the external field, Eq.~(\ref{eq:field}),
one finds:
\begin{equation}
P(t,E) = E\, \Re\left(\chi_1(\omega) \, e^{i \omega t}\right) 
+ \frac{E^3}{4}\Re\left(\chi_{1,2}(\omega) \, e^{i \omega t} + \chi_3 (\omega) \, e^{3 i \omega t} \right) + O (E^5).
\end{equation}
which defines the usual frequency dependent linear susceptibility, $\chi_1(\omega)$, and the frequency dependent non-linear 
susceptibility, $\chi_3 (\omega)$, while $\chi_{1,2}(\omega)$ is the 
$E^2$ correction to the first harmonic susceptibility
$\chi_1(\omega)$. The non-linear susceptibility $\chi_3(\omega)$, which 
is the quantity will focus on throughout this paper, can
be accessed experimentaly by filtering $O(E^3)$ terms at frequency
$3\omega$. 

Following the same procedure, one can expand  in powers of the 
electric field the ($z$-component) polarization correlation and linear response 
functions of the system driven by the electric field $E(t)$.
 The up-down symmetry in the $z$ direction implies that  
they both are even functions of $E$. Therefore their expansion in power of $E$
contains only even terms:
\begin{eqnarray} \label{eq:a2}
\nonumber
C(t,\tpr) &=& C_0 (t, \tpr) +  C_2(t,\tpr) E^2 + O(E^4)\\
R(t,\tpr) &=& R_0 (t, \tpr) +  R_2(t,\tpr) E^2 + O(E^4).
\end{eqnarray}
$C_0$ and $R_0$ are the unperturbed 
correlation and response functions in absence
of the external field. At equilibrium, they are functions only of the time difference 
$\tau = t - \tpr \geq 0$: 
$C_0 (t, \tpr) = C_0 (t - \tpr)$ and 
$R_0 (t, \tpr) = R_0 (t - \tpr)$. 
Moreover, the 
Fluctuation-Dissipation theorem (FDT) holds for the
unperturbed correlation and response functions:
\begin{equation}
R_0 (\tau) = - \, \frac{1}{T} \, \frac{\partial C_0 (\tau)}{\partial \tau}.
\end{equation}
The second-order correlation and response functions
appearing in Eq.~(\ref{eq:a2}) are defined as:
\begin{equation} \label{eq:d2rdef}
C_2(t,\tpr) = \frac{1}{2}
\int_{t_1,t_2 < t} \! \de t_1 \, \de t_2 \, \frac{\delta^2 C
(t,\tpr)}{\delta E (t_1) \, \delta E(t_2)} \bigg | _{E=0} \! \cos (\omega t_1)\,\cos (\omega t_2).
\end{equation}
\begin{equation} \label{eq:d2rrdef}
R_2(t,\tpr) = \frac{1}{2} 
\int_{t_1,t_2<t} \! \! \de t_1 \, \de t_2 \, \frac{\delta^2 R
(t,\tpr)}{\delta E (t_1) \, \delta E(t_2)} \bigg | _{E=0} \! \cos (\omega t_1)\,\cos (\omega t_2).
\end{equation}
The second order correlation function, $C_2(t, \tpr)$, was introduced in the context of spin-glasses
by Huse~\cite{DHuse} in the static limit, and more recently studied in details in~\cite{salerno1,salerno2}.
Neither TTI nor FDT holds for $C_2(t,\tpr)$ and 
$R_2(t,\tpr)$, which are explicit functions of both $t$ and $\tpr$.
However, the response kernels appearing inside the above 
integrals are TTI. Using this property and developing the product of cosines 
in complex exponentials one finds easily that: 
 
\begin{eqnarray} \label{eq:a3}
\nonumber
C_2(t, \tpr) &=& c_0^{(\omega)} (t - \tpr)+ \left(e^{i \omega (t + \tpr)}
\, c_1^{(\omega)} (t - \tpr)+c.c.\right)\\
R_2 (t , \tpr) &=& r_0^{(\omega)} (t - \tpr)+\left(e^{i \omega (t + \tpr)}
\, r_1^{(\omega)} (t - \tpr)+c.c.\right).
\end{eqnarray}

\subsection{Relation between non-linear susceptibility and second-order response}

In the following we aim at establishing a relation between the second
order response function defined above and the non-linear susceptibility.
By definition, the electric polarization is given by the 
convolution of the response function with the external field:
\begin{equation}
P(t) = \int_{-\infty}^t \de \tpr \, R (t, \tpr) \, E(\tpr).
\end{equation}
Therefore, using Eqs.~(\ref{eq:a1}) and (\ref{eq:a2}), we simply get that:
\begin{equation} \label{eq:gad1p}
E P_1(t) =  \int_{-\infty}^t \de \tpr \, R_0 (t - \tpr) \, E(\tpr), 
\end{equation}
and
\begin{equation} \label{eq:gad3p}
E P_3(t) = \int_{-\infty}^t \de \tpr \, R_2(t,\tpr) \, E(\tpr).
\end{equation}
Thus, the component of order $E^3$ of the polarization (related to $\chi_3$) 
turns out to be just the convolution
of the field with the function $R_2(t,\tpr)$ defined in 
Eq.~(\ref{eq:d2rdef}).
As a consequence, using the above expression, together with Eq.~(\ref{eq:a3}), we find, for
an oscillating field at frequency $\omega$:
\begin{equation}
P_3(t) = \Re \Big \{ e^{i \omega t} \left(
 \tilde{r}_0^{(\omega)} (\omega) +  \tilde r_1^{(\omega)} (0) \right ) 
 + e^{3 i \omega t} \tilde{r}_1^{(\omega)} (2 \omega)  
\Big \}\quad,
\end{equation}
where we denoted $\tilde{r}_0^{(\omega)} (\omega')$ and
$\tilde{r}_1^{(\omega)} (\omega')$ the semi-Fourier transform (with respect to $\tau$) at
frequency $\omega'$ of 
the coefficients appearing in the Fourier expansion, $r_0^{(\omega)}(\tau)$ and $r_1^{(\omega)} ( \tau)$. 
The previous equation allows us to establish a general relation between
$\chi_3(\omega)$ and the Fourier transform of $r_{n=1}^{(\omega)} (\tau)$:
\begin{equation} \label{eq:chi3}
\chi_3 (\omega) = 4 \, \tilde{r}_1^{(\omega)} (2 \omega) 
= 4 \int_0^{\infty} \de \tau \, e^{-2 i \omega \tau} \, 
r_{1}^{(\omega)} (\tau).
\end{equation}
This relation will be very useful. Using scaling arguments we will now obtain the  
critical behavior of $r_1^{(\omega)}(\tau)$ within MCT. The 
relation above will then allow us to obtain straightforwardly the 
scaling behaviour of $\chi_3(\omega)$. 

\section{Low frequency limit and relationship to dynamical linear responses}\label{sec:2}

\subsection{Low frequency limit}
In the following we focus on the evolution of the correlation and response
function when the period of oscillation of the external field  is much smaller than 
the relaxation time,  in other words in the low-frequency limit $\omega \tau_\alpha \ll 1$. 
By definition all degrees of freedom relevant for this time-sector of the response or correlation
relax on timescales much smaller than $\omega^{-1}$. As a consequence the correlation/response 
functions are expected to be given by their equilibrium expression in the presence of a quasi-constant
external field $E \cos (\omega t)$. Therefore, in this regime:
\begin{eqnarray} \label{eq:sl}
C(t,\tpr) &=& C_{eq} (t - \tpr, E \cos (\omega t))\\
\nonumber
R(t,\tpr) &=& R_{eq} (t - \tpr, E \cos (\omega t)).
\end{eqnarray}
Since we are interested in the small $E$ behavior, 
we can expand the above expression up to second
order in $E$. For the response function, 
for instance, this yields:
\begin{equation} \label{eq:lfl}
R_{eq}(\tau, E \cos (\omega t)) \approx R_0 (\tau ) + \frac{E^2 \cos^2(\omega t)}{2} \, \frac{\partial^2 R_{eq}(\tau,E)}{\partial E^2} \bigg |_{E=0},
\end{equation}
where $R_0 (\tau)$ is the unperturbed equilibrium response function, 
and the derivative is computed with respect to a constant external field.
Comparing the last equation with Eq.~(\ref{eq:a3}) in the stationary
regime ($t,\tpr \to \infty$ with $\tau = t - \tpr$ finite) we find 
a very simple expression for the $n=1$ component of the expansion $r_1^{(\omega)}(\tau)$ in the regime  $\omega \tau, \omega \tau_\alpha \ll 1$:
\begin{equation}
r_1^{(\omega)} (\tau) = 
\frac{1}{8} \, \frac{\partial^2 R_{eq} 
(\tau,E)}{\partial E^2} \bigg |_{E=0} 
\end{equation}
An analogous relation holds for the correlation function:
\begin{equation}
c_1^{(\omega)} (\tau) = 
\frac{1}{8} \, \frac{\partial^2 C_{eq} (\tau,E)}{\partial E^2} \bigg |_{E=0}
\end{equation}
These general results provide important insights to understand the behaviour of the non-linear susceptibility.
First of all, since the correlation and response functions appearing in 
Eq.~(\ref{eq:sl}) are defined in equilibrium in presence of a constant field,
they must obey FDT. Therefore one can establish a sort of generalized 
Fluctuation-Dissipation relation
between the second order correlation and response functions, which reads:
\begin{equation} \label{eq:fdr}
r_1^{(\omega)} (\tau) = - \, \frac{1}{T} \, \frac{\partial \, c_1^{(\omega)} (\tau)}{\partial \tau}, \end{equation}
which is however only valid in the low frequency domain $\omega \tau, \omega\tau_\alpha \ll 1$. 


\subsection{Relationship with dynamical linear responses}

The results of the previous Section allow us to establish an interesting relation between the second order correlation and response functions
and the {\it dynamical response} $\chi_T(\tau) \equiv T \partial C_{eq}(\tau)/\partial T$ that was recently 
introduced and extensively studied in~\cite{science,JCP1,JCP2,IMCTJ,PRE,Ruocco},
in particular in relation with the behaviour of the four-point dynamical correlation function. The key idea is that in the glassy dynamics 
regime, the equilibrium correlation function $C_{eq}(\tau)$ satisfies to a good approximation the time-temperature superposition (TTS) principle. This 
means that the correlation function for different temperatures, densities, external fields, etc., can be written as a function of 
$\tau/\tau_\alpha(T,\rho,E)$, and the whole $T,\rho,E$ dependence is captured by the structural relaxation time $\tau_\alpha(T,\rho,E)$. 
This becomes actually an exact statement within the $\alpha$-regime of MCT, when the system approaches the dynamical critical point.
In this case, the dynamical critical temperature is expected to show a quadratic 
dependence on the external field (for small fields) of the form:
\begin{equation} \label{eq:tce}
T_{MCT} (E) \approx T_{MCT}(E=0) + \kappa E^2.
\end{equation}
Close to the critical point, a small field changes slightly the 
critical temperature. Since the only thing that matters for the critical 
behaviour is the distance from the critical point, one finds that
applying a small field is equivalent to a small change in temperature. Note that this implies that the relations
found below carry over, within MCT, to the $\beta$-regime as well.

More generally, since $\tau_\alpha(T,\rho,E)$ is expected to be an even function of $E$ because of the up-down symmetry, it should rather
be written as $\tau_\alpha(T,\rho,\Theta)$, with $\Theta=E^2$. Then approximate TTS immediately leads to:
\begin{equation} 
\frac{\partial C_{eq}(\tau)}{\partial \Theta} \approx \frac{\partial \tau_\alpha/\partial \Theta}{\partial \tau_\alpha/\partial T} 
\frac{\partial C_{eq}(\tau)}{\partial T}.
\end{equation}
Around $E=0$ one has $\partial^2 C_{eq}/\partial E^2 = 2 \partial C_{eq}/\partial \Theta$; using the above results 
one finds the very interesting relation (valid for $\omega \tau \ll 1)$:
\begin{equation} \label{eq:chit}
c_1^{(\omega)}(\tau) \approx  \frac{\kappa}{4} \, \frac{\partial C_{eq} (\tau)}{\partial T} = \frac{\kappa}{4 T}\chi_T (\tau),
\end{equation}
where $\kappa=\frac{\partial \tau_\alpha}{\partial \Theta}/\frac{\partial \tau_\alpha}{\partial T}$.
Analogously one finds for the response function (using the FDT relation (\ref{eq:fdr})):
\begin{equation}\label{eqr:chit}
r_1^{(\omega)}(\tau) = - \frac{\kappa}{4 T^2} \, \frac{\partial \chi_T (\tau)}{\partial \tau}.
\end{equation}
Using Eq. (\ref{eq:chi3}) one directly finds the relation between $\chi_3$ and $\widehat \chi_T$ given in Eq. (\ref{eqn:rel}).

One can in fact obtain a slightly more general relation provided the Fourier transform of Eq.~(\ref{eq:chi3}) is dominated by the region $\omega \tau \ll 1$.  One
then finds with some degree of generality that:
\begin{equation} \label{eq:chi3bis}
\chi_3(\omega) \approx  \frac{\partial \chi_1(2 \omega)}{\partial \Theta},
\end{equation}
and finally, using TTS, Eq. (\ref{eqn:rel2}).
This result is important because it establish a firm link with the linear dynamical responses that 
are often used to evaluate dynamical correlations. 

More generally, the amplitude of the correlation function also depends on temperature and electric field:
\begin{equation} 
C_{eq}(\tau) = A(T,E,..) c_{eq}(\frac{t}{\tau_\alpha}),
\end{equation} 
and the derivative of $A$ brings extra contributions that affect the above equalities. In glassy systems, the relaxation time $\tau_\alpha$
is usually most sensitive to external parameters, and it is reasonable to discard these corrections, except at zero frequencies where the 
above contribution is in fact zero. The case in spin-glasses is very different, because the correlation amplitude itself depends critically on temperature. 

\section{Critical behaviour of the
non-linear susceptibility within MCT: scaling arguments} \label{sec:3}

In this section we analyze the behaviour of the non-linear
susceptibility using general physical and scaling arguments. These
results can be confirmed by an exact analysis of the schematic Mode Coupling
(p-spin) equations. For sake of clarity of this paper, 
the technical aspects related to the derivation of the schematic MCT equations in presence of an external
oscillating field and their analysis will presented in a separate publication~\cite{chi3pspin}.

\subsection{Technical preliminaries}

The results above establish a clear connection between the dynamical
response $\chi_T$ and  
the non-linear susceptibility $\chi_3$. In the following, we will
exploit the consequences  
of this connection within the MCT framework, using scaling arguments. 
Note that we will implicitely assume, as previously done in the literature, 
that the results obtained within MCT for density correlation functions
carry out  
to polarization fluctuations. In fact, experimentally, it has been established
that the dielectric susceptibility probes the glassy dynamics as well
as the density correlation  
functions, see e.g.~\cite{NagelMenon}. A quantitative theory of
dielectric polarization fluctuations and  
their coupling to density fluctuations in the slow dynamics regime
would be certainly very involved 
due to the presence of Onsager cavity fields~\cite{Felderhof}.  

Let us now recall the behaviour of $\chi_T(\tau)$ within MCT~\cite{IMCTJ,JCP2}, 
where two critical relaxation regimes occur close to the MCT transition: the $\beta$-regime, with 
relaxation time $\tau_\beta \sim \tau_0 \epsilon^{-1/2a}$, and the $\alpha$-regime, with 
relaxation time $\tau_\alpha \sim \tau_\beta \epsilon^{-1/2b} \gg \tau_\beta$.

The critical properties of $\chi_T(\tau)$ have been derived in terms of scaling functions both in the $\alpha$ and
$\beta$-regimes~\cite{JCP2}. Using these results, the relations established above and 
the FDT of Eq.~(\ref{eq:fdr}), one can obtain the scaling behaviour of $c_1^{(\omega)}(\tau)$ 
and $r_1^{(\omega)}(\tau)$ close to the MCT transition in the different {\em time} regimes. The strategy is the following: 
we start with very low frequencies, where we know the scaling behaviour from the relation with $\chi_T$, and extend it to the whole regimes assuming that critical scaling holds. 
Finally, from the scaling behavior of $c_1^{(\omega)}(\tau)$ 
and $r_1^{(\omega)}(\tau)$ we will obtain the one of $\chi_3(\omega)$.
We will first analyse the $\beta$-regime and afterwards the $\alpha$-regime.

\subsubsection{The $\beta$-regime} \label{sec:2.beta}

Let us first analyze the regime $\omega \tau \ll 1$. In this case, one can use 
the results (\ref{eq:chit},\ref{eqr:chit}) valid in the regime$\tau_0 \ll \tau \sim \tau_\beta \ll \tau_\alpha$ and 
the known behavior of $\chi_T$ ~\cite{IMCTJ,JCP2} to find:
\begin{eqnarray}
c_1^{(\omega)} (\tau) & \approx & \frac{1}{\sqrt{\epsilon}} \, c_\beta\left(\frac{\tau}{\tau_{\beta}} \right)\\
\nonumber
r_1^{(\omega)} (\tau) & \approx & \frac{1}{\sqrt{\epsilon}} \, \frac{1}{\tau_{\beta}} \, r_\beta \left( \frac {\tau}{\tau_{\beta}} \right),
\end{eqnarray}
where the scaling function $c_\beta (x)$ behaves asymptotically as $x^a$ for $x \ll 1$ and as $x^b$ for $x \gg 1$, whereas $r_\beta (x)$ 
behaves as $x^{a-1}$ for $x \ll 1$ and as $x^{b-1}$ for $x \gg 1$.

On the other hand, for very large frequencies, such that $\omega \tau \gg 1$,
the field oscillates so fast that the system has no time
to respond and one expects vanishing (for $\omega\tau \rightarrow \infty$) 
small second order correlation and response
functions. 

Finally, for $\omega \tau$ of the order of one,  the most general scaling behavior  
in the $\beta$-regime generalizing the one above reads:
\begin{eqnarray} \label{eq:betageneral}
c_1^{(\omega)}(\tau) & \approx & \frac{1}{\sqrt{\epsilon}} \, \hat f_{\beta} \left( \frac{\tau}{\tau_{\beta}},\omega \tau \right) \\
r_1^{(\omega)}(\tau) & \approx & \frac{1}{\sqrt{\epsilon}} \, \frac{1}{\tau_{\beta}} \,\hat g_{\beta} \left( \frac{\tau}{\tau_{\beta}},\omega \tau\right).
\end{eqnarray}
The most general assumption compatible with previous results is a factorized form for $\hat f_\beta$ and $\hat g_\beta$ in both regimes
$\tau \gg \tau_\beta$ and $\tau \ll \tau_\beta$. In the late $\beta$-regime, one has (with $L$ for `late'):
\begin{eqnarray}\label{asymp:betasmall1}
c_1^{(\omega)} (\tau) & \simeq & \frac{1}{\sqrt{\epsilon}} \, \left( \frac{\tau}{\tau_{\beta}}\right)^b 
\ofb(\omega \tau) \\\label{asymp:betasmall2}
r_1^{(\omega)} (\tau) & \simeq & \frac{1}{\sqrt{\epsilon}} \, \frac{1}{\tau_{\beta}} \,
\left( \frac{\tau}{\tau_{\beta}} \right)^{b-1} \ogb(\omega \tau).
\end{eqnarray}
where both functions $\ofb, \ogb$ tend to a constant when their argument $\omega \tau$ is small, and tend to zero
when $\omega \tau$ is large. In the early $\beta$-regime, a similar result holds, with a priori different scaling functions
$f^E_\beta, g^E_\beta$ ($E$ for `early'):
\begin{eqnarray}\label{asymp:betalarge1}
c_1^{(\omega)} (\tau) & \simeq & \frac{1}{\sqrt{\epsilon}} \, 
 \left( \frac{\tau}{\tau_{\beta}}\right)^a f^E_\beta(\omega \tau) \\\label{asymp:betalarge2}
r_1^{(\omega)} (\tau) & \simeq & \frac{1}{\sqrt{\epsilon}} \, \frac{1}
{\tau_{\beta}} \,
\left( \frac{\tau}{\tau_{\beta}} \right)^{a-1} g^E_\beta(\omega \tau).
\end{eqnarray}
These new functions $f^E_\beta, g^E_\beta$ again  tend to a constant when their argument $\omega \tau$ is small, and tend to zero
when $\omega \tau$ is large. 

In this regime, the explicit dependence with $\tau$ and $\omega$
occurs only through the rescaled time and frequency $\tau/\tau_\beta$
and $\omega \tau_\beta$. In the rest of the text, we will frequently
use the variables $\hat\tau=\tau/\tau_\beta$, $\hat\omega=\omega
\tau_\beta$ and $x=\omega\tau$.

\subsubsection{The $\alpha$-regime} \label{sec:2.alpha}

The behaviour of $c_1^{(\omega)},r_1^{(\omega)}$ in the $\alpha$-regime follows similar scaling laws.
When $\omega\tau \ll 1$, the results of ~\cite{IMCTJ,JCP2} allow one to obtain:
\begin{eqnarray}
c_1^{(\omega)} (\tau) & \approx & \frac{1}{\epsilon} \, c_\alpha \left(\frac
{\tau}{\tau_{\alpha}} \right)\\
\nonumber
r_1^{(\omega)} (\tau) & \approx & \frac{1}{\epsilon} \, \frac{1}
{\tau_{\alpha}} \, r_\alpha \left( \frac {\tau}{\tau_{\alpha}} \right),
\end{eqnarray}
The matching between the late $\beta$-regime and the $\alpha$-regime
determine the asymptotic behaviour of the scaling functions defined above.
One finds that $c_\alpha(x \ll 1)$ behaves as $x^b$ and $r_\alpha (x \ll 1)$ as $x^{b-1}$, 
whereas both functions tend exponentially fast to zero for $x \gg 1$.

When $\omega \tau$ is not small, the scaling behaviour in the
$\alpha$-regime reads:  
\begin{eqnarray} \label{asymp:alpha2_1}
c_1^{(\omega)}(\tau) & \approx & \frac{1}{\epsilon} \, \overline f_{\alpha} \left( \frac{\tau}{\tau_{\alpha}},\omega \tau \right) \\\label{asymp:alpha2_2} 
r_1^{(\omega)}(\tau) & \approx & \frac{1}{\epsilon} \, \frac{1}{\tau_{\alpha}} \,\overline g_{\alpha} \left( \frac{\tau}{\tau_{\alpha}},\omega \tau\right).
\end{eqnarray}
In the early $\alpha$-regime, that is when $\tau \ll \tau_\alpha$, one finds:
\begin{eqnarray}\label{asymp:alphalarge1}
c_1^{(\omega)} (\tau) & \simeq & \frac{1}{\epsilon} \, 
\left( \frac{\tau}{\tau_{\alpha}} \right )^b f_{\alpha}^E 
(\omega \tau) \\\label{asymp:alphalarge2}
r_1^{(\omega)} (\tau) & \simeq & \frac{1}{\epsilon} \, \frac{1}
{\tau_{\alpha}} \,
 \left( \frac{\tau}{\tau_{\alpha}} \right )^{b-1} g_{\alpha}^E (\omega \tau).
\end{eqnarray}
Furthermore, by requiring the matching 
between the two regimes of large $\tau/\tau_\beta$  and small $\tau/\tau_\alpha$, one finds that the scaling functions 
$f_{\alpha}^E (x)$ and $g_{\alpha}^E (x)$ 
are the same as $f^L_{\beta} (x)$ and $g^L_{\beta} (x)$.

As in the $\beta$-regime, the explicit dependence with $\tau$ and $\omega$
occurs only through the rescaled time and frequency $\tau/\tau_\alpha$
and $\omega \tau_\alpha$. In the rest of the text, we will frequently
use the variables $\overline\tau=\tau/\tau_\alpha$, $\overline\omega=\omega
\tau_\alpha$ (and  also $x=\omega\tau$).

\begin{figure}
\begin{center}
\includegraphics[width=0.64\textwidth]{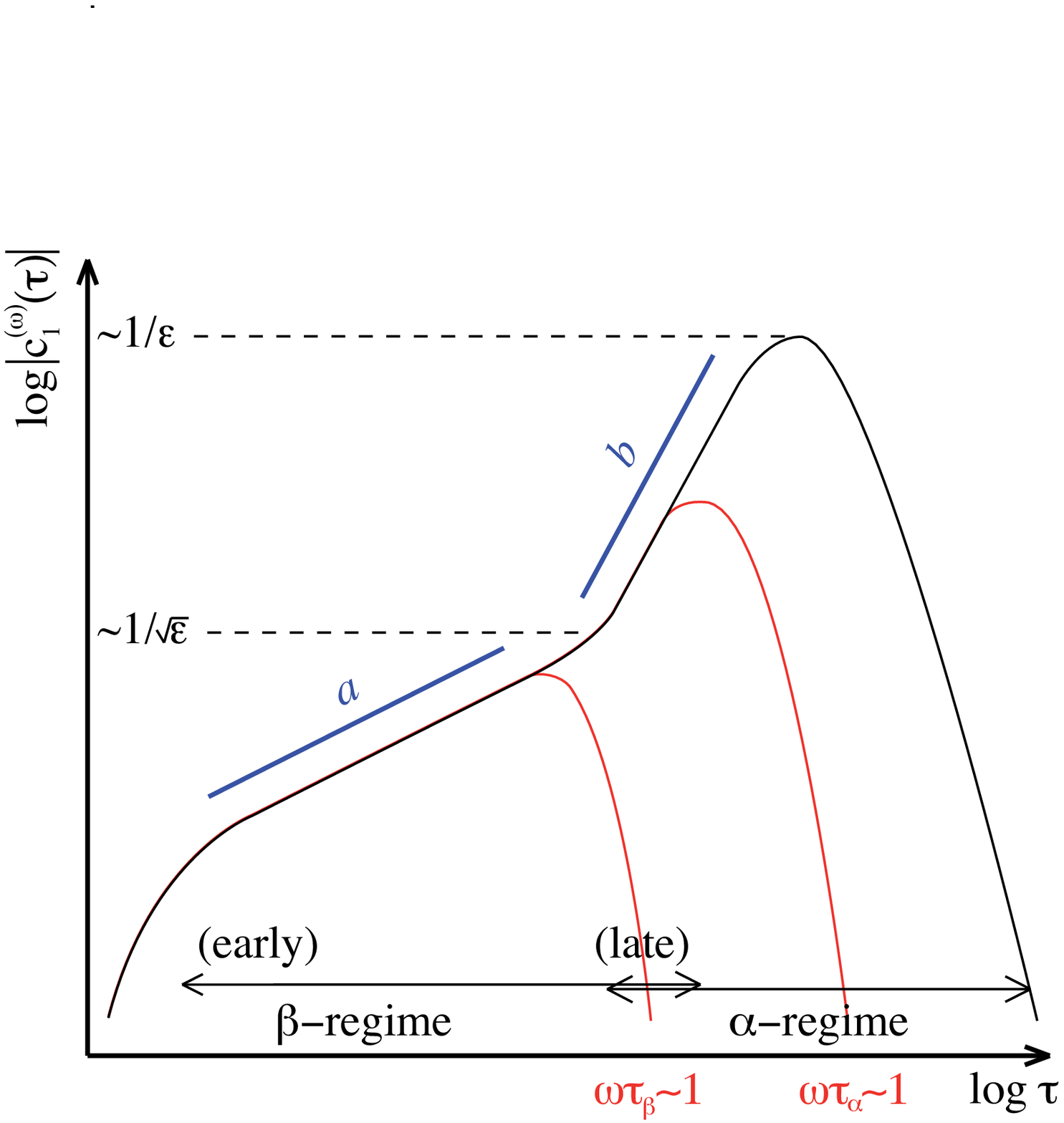}
\end{center}
\caption{Sketch of $\log |\dco (\tau)|$ as a function of $\log \tau$.
In the $\omega \to 0$ limit $\dco (\tau)$, behaves as $\chi_T (\tau)$,
i.e., it scales as $1/\sqrt{\epsilon} \, (\tau/\tau_{\beta})^a$ in the early
$\beta$-regime and as $1/\sqrt{\epsilon} \, (\tau/\tau_{\beta})^b$ in the late
$\beta$-regime (or, equivalently, as $1/\epsilon (\tau/\tau_{\alpha})^b$ in the
early $\alpha$-regime). In the $\alpha$-regime $|\dco (\tau)|$ reaches 
a maximum of order $1/\epsilon$. At finite frequency $\omega$, $|\dco
(\tau)|$ drops to values of $O(1)$ as $\tau \gtrsim 1/\omega$ (red curves).}
\label{fig:dc1}
\end{figure}

In Fig.~\ref{fig:dc1} we show a sketch of the behavior of
$|c_1^{(\omega)}(\tau)|$ that summarizes all our previous findings.  

\subsection{Scaling behavior of  $\chi_3 (\omega)$}

In the previous section we have determined 
the scaling forms governing the time and temperature dependence of
$r_1^{(\omega)}(\tau)$ and $c_1^{(\omega)}(\tau)$. 
Using these results we can now easily analyze the critical behaviour of the non-linear susceptibility, by computing its 
Fourier transform at frequency $2 \omega$, according to Eq.~(\ref{eq:chi3}). We again focus in turn on the $\beta$-regime 
and then on the $\alpha$-regime, before commenting on the zero and infinite frequency limits.

\subsubsection{The $\beta$-regime}

Let us first consider probing frequencies of the order of the inverse
of the $\beta$-relaxation time. 
We set $\hat{\omega}= \omega \tau_{\beta}$ and assume that only the
$\beta$-regime of $r_1^{(\omega)}(\tau)$ contributes to
Eq.~(\ref{eq:chi3}). This assumption can be fully justified by the
exact analysis of the 
schematic Mode Coupling equations~\cite{chi3pspin}. 
Indeed one can show that  
$r_1^{(\omega)}(\tau \sim \tau_{\alpha})$ is small 
due to the fact that 
the scaling function $g_{\alpha}^E (x)$, introduced in 
Eq.~(\ref{asymp:alphalarge2}), vanishes as $1/x^{1+b}$ at large $x$.
Therefore, for probing frequencies of the order of the inverse of the
$\beta$-relaxation time, the contribution to the non-linear susceptibility 
coming from the time integral in the $\alpha$-regime is negligible in Eq.~(\ref{eq:chi3}).
 
One then finds that:
\begin{equation}\label{scalingbeta}
\chi_3 (\omega) \simeq \frac{1}{\sqrt{\epsilon}} \, {\cal F} (\hat{\omega})
\end{equation}
where the function ${\cal F}(x)$ is defined as:
\[
{\cal F}(\hat{\omega})=\frac{4}{\hat \omega} \int_0^\infty \de u \,
e^{-2iu} \, \hat g_\beta \left(\frac{u}{\hat \omega},u \right) 
\]
The asymptotic behaviour of the scaling function ${\cal F}$ can easily be obtained from the results of the previous section. One
finds:
\begin{eqnarray}
{\cal F}(\hat{\omega}) & \simeq &  4{\hat \omega}^{-b} \int_0^\infty \de u \, u^{b-1} e^{-2iu} g^L_\beta (u) \qquad \omega \tau_\beta \ll 1\\
                       & \simeq &  4{\hat \omega}^{-a} \int_0^\infty \de u \, u^{a-1} e^{-2iu} g^E_\beta (u) \qquad \omega \tau_\beta \gg 1.           
\end{eqnarray}
Using the asymptotic properties of $g^L_\beta (u), g^E_\beta (u)$ and the fact that $a,b$ are between zero and one 
insures the convergence of the integrals appearing in the above
equation, at both small and large $u$. This confirms that 
the scaling behaviour of $\chi_3(\omega)$ in this region is indeed dominated by the $\beta$-regime of $r_1^{(\omega)}(\tau)$. 
Note that in the high frequency region the $\epsilon$ dependence of
$\chi_3(\omega)$ drops out, as it should in order to match the
non critical $\tau_0^{-1}$ frequency regime. 

\subsubsection{The $\alpha$-regime}

We now consider the $\alpha$-regime, where we set $\overline{\omega}=
\omega \tau_{\alpha}$, and again assume that only  
the scaling form of $r_1^{(\omega)} (\tau)$ in this same regime, Eq.~(\ref{asymp:alpha2_2}), contributes significantly to Eq.~(\ref{eq:chi3}). 
Indeed, the contribution due to the time integral in the 
$\beta$-regime is at least 
a factor $\sqrt{\epsilon}$ smaller than the
one coming from the $\alpha$-regime, and yields a subleading contribution to the critical behavior 
of $\chi_3 (\omega)$.
We then
find that the non-linear susceptibility scales as:
\begin{equation}\label{scalingalpha}
\chi_3 (\omega) \simeq \frac{1}{\epsilon} \, {\cal G} (\overline{\omega})
\end{equation}
where the function ${\cal G}(x)$ is defined as:
\[
{\cal G}(\overline{\omega})=\frac{4}{\overline \omega}\int_0^\infty
\de u \, e^{-2iu} \, \overline g_\alpha^E \left(\frac{u}{\overline \omega},u
\right) 
\]
Clearly, because of the matching of $\overline g_\alpha$ for small
first arguments with $\hat g_\beta$ at large first arguments,
we find that the scaling of the early $\alpha$-regime ($\omega \tau_\alpha \gg 1$) of the non-linear susceptibility matches with that of 
the late $\beta$-regime ($\omega \tau_\beta \ll 1$), with:
\begin{equation}
\chi_3(\omega) \propto \epsilon^{(b-a)/2a} \omega^{-b}.
\end{equation}
 
\subsubsection{Low frequency limit}

In the low frequency limit, one finds that $\chi_3(\omega)$ decreases from its peak value $\epsilon^{-1}$ reached for 
$\omega \sim 1/2\tau_\alpha$ to a non critical, finite value given by Eq.~(\ref{eq:chi3bis}). 
As discussed in~\cite{BB-PRB}, contrary to the case of spin-glasses,
the non-linear susceptibility is critical only for small but non zero values of the frequencies. Zero frequency
corresponds to a static equilibrium response (or correlation, via FDT). In glasses, these are not expected to have 
any critical behavior.

In particular, in the low frequency limit, one can expand Eq.~(\ref{eq:chi3}) up to second order in
$\omega\tau$. Using Eq.~(\ref{eq:lfl}) we have:
\begin{eqnarray}
\chi_3 (\omega) &\approx& \kappa \int_0^{\infty} \de \tau \left ( 1 - 2 i \omega \tau - 2 \omega^2
\tau^2 \right) \frac{\partial R_{eq} (\tau)}{\partial T } \\
\nonumber
&\approx& \kappa \frac{\de }{\de T}  \{ \chi_1 (0) \left (1 - 2 i \omega A_1 \tau_{\alpha} - 4 \omega^2 A_2 \tau_{\alpha}^2 \right ) \},
\end{eqnarray}
where the zero frequency limit of the linear susceptibility, $\chi_1 (0)$, equals the static polarization fluctuations (along the z-axis) 
divided by temperature, $N\langle P^2\rangle/T$. $A_1$ and $A_2$ are two temperature-independent constants defined as
$A_i = \int_0^{\infty} ds s^{i} \, r_{eq} (s)$
(here we have used again the time-temperature superposition principle, writing $R_{eq}(\tau)=\chi_1 (0) r_{eq}(\tau/\tau_\alpha)/\tau_\alpha$).
The last equations allows us to determine the low frequency behavior of the real and imaginary
part of the non-linear susceptibility:
\begin{eqnarray}
\nonumber
\Re \left ( \chi_3 (\omega) \right ) &\approx& \kappa \frac{\de 
\chi_1(0)}{\de T} + B_1 \omega^2 + O ( (\omega\tau_\alpha)^4 ) \\
\nonumber
\Im \left ( \chi_3 (\omega) \right ) &\approx& B_2 \omega + O (( \omega\tau_\alpha)^3 ),
\end{eqnarray}
with $B_1 = - 4 \kappa A_2 \omega^2 \, \de(\chi_1 (0)  \tau_{\alpha}^2 )/ \de T > 0$ and
$B_2 =- 2 \kappa A_1 \omega   \, \de (\chi_1 (0)  \tau_{\alpha})/ \de T > 0$.

\subsubsection{Large frequency limit}

At very large frequencies (very small timescales) the non-linear
susceptibility is vanishing because the system has not enough time to respond to the oscillating field. 
One could argue that the analysis of Eq.~(\ref{eq:d3p}) at very large frequencies yields:
\begin{equation}
P_3(t) \sim \frac{\delta^3 P}{\delta E^3 (0)} \bigg |_{E=0} 
\left[ \int_0^t \de 
t_1 \, (e^{i \omega t_1} + e^{- i \omega t_1} ) \right]^3.
\end{equation}
As a result, at very large frequency the non-linear susceptibility behaves as:
\begin{equation}
\chi_3 (\omega \to \infty) \sim \frac{1}{(i \omega)^3} \, 
\frac{\delta^3 P}{\delta E^3 (0)} \bigg |_{E=0}.
\end{equation}
This analysis is oversimplified and assumes analytic properties of the function 
$\frac{\delta^3 P}{\delta E(t_1)\delta E(t_2)\delta E(t_3)}$ that are not granted and may depend 
strongly on the microscopic dynamics. For instance, in the case of Ising spins with a Monte Carlo heath bath
dynamics one can easily verify that the previous arguments do not apply and the large frequency 
behavior is proportional to $1/(i\omega)$. The conclusion is that the high frequency behavior depends 
on the underlying microscopic dynamics and, likely, on the type of the non-linear response considered. 
In the case of non-linear dielectric susceptibility the underlying microscopic dynamics should be provided
by Langevin equations for dipoles in a non-polar solvent (this is an approximation since at extremely high 
frequency inertia effects will play a role). To work out the high frequency behavior one can neglect 
interactions with other dipoles and the coupling to structural relaxation. Thus, 
the analysis of the non-linear response of a single dipole in a non-polar solvent worked out in~\cite{Dejardin} 
should apply. The outcome is the $1/(i\omega)^3$ behavior discussed above.           

\subsection{Beyond schematic Mode Coupling: general considerations and non-linear responses to time and space inhomogeneous fields}

The previous scaling arguments are rather general. We emphasis here that they can be fully derived from a rigourous analysis of schematic Mode Coupling equations, in the framework of the spherical p-spin model in presence of an oscillating external field. Analysing the equations up to second order in the external field, it is possible to determine the critical behaviour obtained using scaling arguments in the previous sections.

Beyond the schematic approach, one may wonder about specific but important details such as the role
of conserved variables like energy or density on the above results. As recalled in the introduction, we know that these
conserved variables can dramatically change the scaling behaviour of $\chi_4$ for example, which also diverges as
$(T-T_c)^{-1}$ within a p-spin framework with Langevin dynamics, but diverges as $(T-T_c)^{-2}$ when the contribution
of conserved variables is taken into account~\cite{JCP1,JCP2}. From a diagrammatic point of view, this is due to the
presence of `squared ladder' diagrams which gives the dominant contribution to $\chi_4$. One can check that due
to the causality of the response functions, these diagrams in fact are absent when one computes the non-linear susceptibility
and the above results are expected to hold for a {\it bona fide} MCT theory of liquids. Thus, the result~\cite{BB-PRB}:
\begin{equation}
\chi_3 (\omega) \simeq \xi^{2 - {\overline \eta}} \, {\cal G} (\omega \tau_\alpha),
\end{equation}
is expected to hold even beyond schematic MCT ($\xi$ is the dynamical correlation length which also appears in $\chi_T$). A complete proof could be obtained generalizing the Inhomogeneous MCT calculation of~\cite{IMCTJ} to account for a space and time dependent source term, that would 
describe the non-linear response to an oscillating field with wave-vector $q$ and frequency $\omega$. In a first attempt, we adapt the previous scaling arguments to the explicit results of~\cite{IMCTJ} on the
wavevector-dependent 
dynamical response. This allow us to obtain the critical behavior of 
$\chi_3(\omega,q)$. In the $\beta$-regime one finds:
\[
\chi_3(\omega,q)=\xi^2 H_{\beta}(\omega \tau_{\beta},q\xi) \quad
\xi=\epsilon ^{-1/4}, \quad \tau_{\beta}=\epsilon ^{-\frac{1}{2a}} 
\] 
where the scaling function $H_\beta(x,y)$ is equal to ${\cal F}(x)$
for $y=0$, i.e. for a uniform electric field one finds back
Eq.~(\ref{scalingbeta}). For large $y$ one expects a power law
behavior such as $\frac{h_\beta(x)}{y^2}$ (where $h_\beta(x)$ is a
certain scaling function). As discussed in~\cite{IMCTJ} this is needed
to cancel out the diverging prefactor $\xi^2$ and match the critical
behavior to the non-critical one taking place for $q\propto O(1)$. The
asymptotic behavior with respect to $x$ is identical 
to the one already described for homogeneous fields. Very small $x$ correspond to the matching between $\alpha$ and $\beta$ regimes. 
Since in the $\alpha$ regime $\chi_3$ is expected to diverge as
$\xi^4$, the matching imposes the behavior at small $x$:
$\frac{h^L_\beta(y)}{x^b}$  (where $h^L_\beta(y)$ is another scaling
function). For large $x$ values, the field varies so rapidly that the
system has not enough time to adjust and to respond to the
field. Again, in order to cancel the diverging prefactor and match the
non-critical behavior one expects a large $x$ behavior such as
$\frac{h^E_\beta(y)}{x^a}$  (where $h^E_\beta(y)$ is a third scaling
function). It would be interesting to specify in more details the
shape of the 
scaling functions $h_\beta, h^L_\beta, h^E_\beta$. 

In the $\alpha$ regime one expects: 
\[
\chi_3(\omega,q)=\xi^4 H_{\alpha}(\omega \tau_{\alpha},q\xi) \quad \xi=\epsilon ^{-1/4}, \quad \tau_{\alpha}=\epsilon ^{-\frac{1}{2a}-\frac{1}{2b}}
\]
where the scaling function $H_\alpha(x,y)$ is equal to ${\cal G} (x)$ for $y=0$, i.e. for a uniform electric field one finds back eq. (\ref{scalingalpha}). The same kind of arguments used above suggests for large $y$ a power law behavior: $H_{\alpha}(x,y)\simeq \frac{h_\alpha(x)}{y^4}$ (where $h_\alpha(x)$ is a scaling function). For very small $x$ the scaling function vanishes in order to match the $x=0$ value corresponds to the non-critical (non diverging) static non-linear susceptibility. For large $x$ values in order to match the $\beta$ regime one expects a 
behaviour such as $\frac{h^E_\alpha(y)}{x^b}$  (where $h^E_\alpha(x)$ is another scaling function).

At this point, one may debate about the validity of the power law divergence of the relaxation time described by Mode Coupling Theory and of TTS. However, it is often observed that there are actually regimes, when the dynamics starts slowing down strongly, which are well described by MCT (and thus TTS). In these regimes we expect our predictions for $\chi_3$ to hold, as it seems to be the case for $\chi_4$~\cite{Andersen}. Furthermore, it was recently shown that MCT may be corrected including higher order terms. It was found that these corrections only affect the values of exponents $a$ and $b$, and $T_c$, but do not affect strongly scaling functions~\cite{Andreanov,Mayer}. This suggest that the MCT mechanism for describing the slowing down is rather robust and that the MCT regime could in principle being expanded at the price of changing the exponents (in particular the one coverning the relaxation time, see e.g.~\cite{Mayer})

\section{Conclusion}\label{sec:concl}

In this work, we have studied in detail the non-linear response of supercooled liquids. Although we are able
to provide precise statements within a Mode-Coupling approach, some of our results are in fact more general 
and only require Time-Temperature Superposition to hold. In particular we expect our results to hold in generalization 
of Mode Coupling Theory such as~\cite{Bagchi}.

An important theoretical result is the relation (\ref{eq:chi3bis})
between the non-linear response $\chi_3(\omega)$ and the temperature derivative of the usual linear susceptibility, $d\chi_1(2\omega)/dT$,
valid at small frequencies (smaller than the inverse relaxation time). This bridges the gap between non-linear responses and probes of dynamic
correlations such as three- and four-point correlations and temperature (or density) derivative of standard two-body correlations and
response: they are all different facets of the same underlying physical phenomenon.  

For larger frequencies - of the order of the inverse of the relaxation time - we have obtained 
a complete set of scaling predictions concerning the critical behavior of $\chi_3(\omega)$ 
within MCT. The main results are summarized in Fig.~\ref{fig:chi3}. 
Five different frequency regimes are identified: $\chi_3 (\omega)$ exhibits a peak around
frequencies of the order of half the inverse of the structural relaxation
time of the system $\tau_\alpha$. The height of the peak grows as $(T-T_c)^{-1}$ (or equivalently as $\xi^4$) as the 
critical temperature is approached. For higher frequencies, $\chi_3(\omega)$ decays as power laws, with an exponent equal
to $-b$ in the late $\beta$-regime, to $-a$ in the early $\beta$-regime, and finally to $-3$ at high frequencies. 

Our results should be directly applicable to the non-linear dielectric constant of molecular glasses in the weakly supercooled
regime where MCT is expected to be relevant, and for describing the non-linear compressibility 
or more general non-linear rheological responses
of hard-sphere colloids close to the glass transition, where MCT does a fair job at describing their relaxation properties. However, it is well known that MCT
fails for deeply supercooled liquids, when activated events start playing a major role in the relaxation. The detailed shape of 
$\chi_3(\omega)$ would require a full theoretical description of the dynamics in this regime, which is unavailable to date. 
Still, the general low frequency relation between $\chi_3$ and $d\chi_1/dT$, supplemented with the property of Time-Temperature
superposition, allows one to give a firmer basis to the scaling relation conjectured in~\cite{BB-PRB}, namely that:
\begin{equation}
\chi_3 (\omega) \approx \chi_3^* \, {\cal G} (\omega \tau_\alpha),
\end{equation}
where ${\cal G}$ is a scaling function, and $\chi_3^* \propto d\ln \tau_\alpha/d\ln T$ is the peak value of the temperature derivative of 
$\chi_1(\omega)$, as measured in~\cite{science,PRE,Ruocco}. Following~\cite{science,JCP1,JCP2}, we expect $\chi_T^*$ to increase  as a power of the
dynamical correlation length $\xi(T)$. The detailed shape of ${\cal G}$ would obviously be worth knowing in order to 
compare with upcoming experimental results. As a guide, we give the result obtained assuming a Havriliak-Nagami form for the 
susceptibility and the validity of the relation between $\chi_3$ and $d\chi_1/dT$ at all frequencies, which has no justification
apart from suggesting possible fitting functions. One finds:
\begin{equation}
{\cal G}_{HN} (u) = \frac{(iu)^b}{(1 + (iu)^b)^{1+c}} ,
\end{equation}
where $b,c$ are fitting exponents.

Among open problems worth investing is the extension of the present theory to the aging regime of glasses and spin-glasses. 
From an experimental point of view, a detailed study of the role of the electric field on the glass properties of dipolar
liquids (such as glycerol) would be very interesting. For example, the evolution of the glass transition temperature as 
a function of the field $E$ would allow one to measure the proportionality coefficient $\kappa$ appearing in Eq.~(\ref{eq:chi3bis}).
In spin-glasses, a detailed measurement of $\chi_3 (\omega)$ would allow to shed light on the existence of spin-glass transition at non 
zero field, as argued in~\cite{BB-PRB}. The behaviour of $\chi_3 (\omega,t_w)$ in the aging phase would furthermore be a very useful 
probe of the aging process in spin-glasses, in particular during rejuvenation cycles. Numerical simulations 
of  $\chi_3 (\omega,t_w)$, using the zero-field techniques developed in~\cite{salerno1,salerno2}, would be 
worth pursuing.

\section*{Acknowledgements} We want to thank F. Ladieu, D. L'H{\^o}te and C. Thibierge for motivating this study and for useful discussions
on their experimental results. We also thank L. Berthier, A. Billoire, F. Corberi, M. Fuchs, K. Miyazaki, R. Richert, D. Reichman, T. Sarlat, H. Yoshino \& F. Zamponi for interesting remarks.  

\section*{List of figures}

{\bf Figure 1}
Sketch of $\log |\chi_3 (\omega)|$ as a function of
$\log \omega$, showing five different frequency regimes: $\omega
\tau_\alpha\ll 1$, $\omega \tau_\alpha \sim 1$,
$\tau_\beta/\tau_\alpha\ll\omega \tau_\beta \ll 1$, $\omega \tau_\beta \gg 1$, $\omega \tau_0 \sim 1$.
Note that the low frequency limit is non zero but much smaller than the peak value for $T$ close to $T_c$.\\\\
{\bf Figure 2}
Sketch of $\log |\dco (\tau)|$ as a function of $\log \tau$.
In the $\omega \to 0$ limit $\dco (\tau)$, behaves as $\chi_T (\tau)$,
i.e., it scales as $1/\sqrt{\epsilon} \, (\tau/\tau_{\beta})^a$ in the early
$\beta$-regime and as $1/\sqrt{\epsilon} \, (\tau/\tau_{\beta})^b$ in the late
$\beta$-regime (or, equivalently, as $1/\epsilon (\tau/\tau_{\alpha})^b$ in the
early $\alpha$-regime). In the $\alpha$-regime $|\dco (\tau)|$ reaches 
a maximum of order $1/\epsilon$. At finite frequency $\omega$, $|\dco
(\tau)|$ drops to values of $O(1)$ as $\tau \gtrsim 1/\omega$ (red curves).

\end{document}